\documentclass[aip,apl,reprint,superscriptaddress,twocolumn,showpacs]{revtex4-1}

\usepackage[none]{hyphenat}
\usepackage{graphicx}
\usepackage{dcolumn}
\usepackage{bm}
\usepackage{amsmath}
\usepackage{color}
\usepackage[colorlinks,
linkcolor=blue,
citecolor=blue,
urlcolor=blue]{hyperref}

\begin{document}

\title{Magnetic field amplification to the gigagauss scale via dynamos driven by femtosecond lasers}

\author{K. Jiang} \email{jiangke18@gscaep.ac.cn}
\affiliation{Institut f\"{u}r Theoretische Physik I, Heinrich-Heine-Universit\"{a}t D\"{u}sseldorf, 40225 D\"{u}sseldorf, Germany}
\affiliation{Graduate School, China Academy of Engineering Physics, Beijing 100088, China}

\author{A. Pukhov} \email{pukhov@tp1.uni-duesseldorf.de}
\affiliation{Institut f\"{u}r Theoretische Physik I, Heinrich-Heine-Universit\"{a}t D\"{u}sseldorf, 40225 D\"{u}sseldorf, Germany}

\author{C. T. Zhou} \email{zcangtao@sztu.edu.cn}
\affiliation{Center for Advanced Material Diagnostic Technology, and College of Engineering Physics, Shenzhen Technology University, Shenzhen 518118, China}

\date{\today}

\begin{abstract}
Reaching gigagauss magnetic fields opens new horizons both in atomic and plasma physics. At these magnetic field strengths, the electron cyclotron energy $\hbar\omega_{c}$ becomes comparable to the atomic binding energy (the Rydberg), and the cyclotron frequency $\omega_{c}$ approaches the plasma frequency at solid state densities that significantly modifies optical properties of the target. The generation of such strong quasistatic magnetic fields in laboratory remains a challenge. Using supercomputer simulations, we demonstrate how it can be achieved all-optically by irradiating a micro-channel target by a circularly polarized relativistic femtosecond laser. The laser pulse drives a strong electron vortex along the channel wall, inducing a megagauss longitudinal magnetic field in the channel by the inverse Faraday effect. This seed field is then amplified up to a gigagauss level and maintained on a picosecond time scale via dynamos driven by plasma thermal expansion off the channel walls. Our scheme sets a possible platform for producing long living extreme magnetic fields in laboratories using readily available lasers. The concept might also be relevant for applications such as magneto-inertial fusion. 
\end{abstract}
\maketitle
\sloppy{}
The study of matter in magnetic fields of different scales has been an important research subject since the discovery of magnetic field for hundreds of years. The modern science pursues a number of phenomena appearing in strong magnetic fields, such as the quantum hall effect in condensed matter \cite{Zhang}, the Zeeman effect in atoms \cite{Zeeman}, macro- \cite{Li} and microscope instabilities \cite{Bret} in plasma, etc. The imposed magnetic fields currently available in laboratories are at megagauss (MG) levels or below. At the same time, astrophysics objects possess magnetic fields with amplitudes within a huge range from microgausses to teragausses and even beyond. When the magnetic field strength reaches gigagauss (GG) level, distinct new physics comes forth. In such extreme environment, the magnetic cyclotron energy ($\mathcal{E}_{c}=\hbar eB/m_{e}c$) of an electron becomes comparable to its atomic Coulomb binding energy (the Rydberg $R_{y}=\alpha^{2}m_{e}c^{2}/2$, where $\alpha=e^{2}/\hbar c$ is the fine structure constant). The Zeeman effect becomes quadratic for atomic hydrogen, leading to quantum chaos \cite{Zhou,Friedrich}, which is evidenced by unexpected astrophysical spectra from high-field magnetic white dwarfs \cite{Vanlandingham}. The quadratic Zeeman effect is currently under active investigations in atomic physics and astrophysics. Beyond this threshold, electrons follow the Landau quantization in ultra-strong magnetic fields instead of the Bohr orbits \cite{Landau}. On the other hand, in the context of high energy density physics \cite{Davidson}, the electron cyclotron frequency ($\omega_{c}=eB/m_{e}c$) approaches the plasma frequency ($\omega_{pe}=\sqrt{4\pi n_{e}e^{2}/m_{e}}$) at solid state densities for magnetic fields at the GG scale. The plasma becomes strongly magnetized, and all its properties are dominated by the influence of the imposed magnetic field. It is theoretically predicted that in such a regime, right-handed circularly polarized (CP) lasers can penetrate freely along the magnetic force lines into dense plasma without cutoff or resonance \cite{Chen}. This sheds new light on optical diagnosis of opaque materials and magnetically assisted fast ignition \cite{Wang}. 

So far, magnetic fields at GG scale are still inaccessible in laboratories, thus investigations in aforementioned areas have been limited to theoretical studies \cite{Ma} and laboratory analogues \cite{Murdin}. At the same time, the production of GG-scale magnetic fields is essential for further understanding the new effects at extreme magnetic environments in wide areas such as atomic physics, high energy density physics, astrophysics, etc. It might also be relevant for applications such as magneto-inertial fusion \cite{Basko,Lindman} and charged beam manipulation \cite{Zhuo}.

\begin{figure*}
	\centering
	\includegraphics[width=17.2cm]{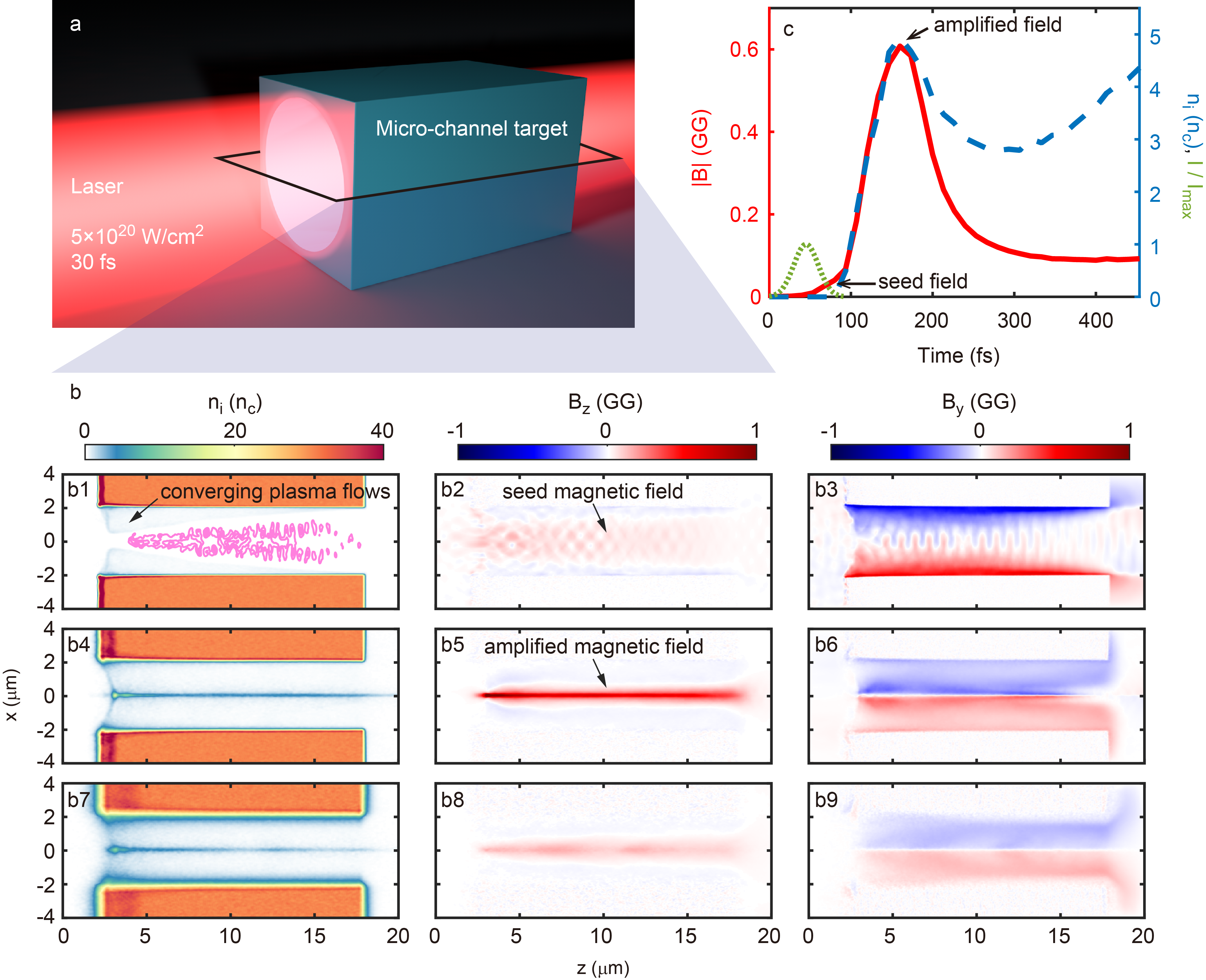}
	\caption{{\bf Schematic diagram and 3D-PIC simulation results} (\textbf{a}) A circularly polarized laser pulse with wavelength $\lambda=800$ nm, peak intensity $I_0=5\times10^{20}$ W/cm$^{2}$ and duration $\tau=30$ fs irradiates a carbon micro-channel target. The 16 $\mu$m long target with edge width of 8 $\mu$m and vacuum hollow radius of 2 $\mu$m is aligned with the laser propagation direction. (\textbf{b}) Simulation results showing the converging plasma flows and associated magnetic field amplification. (b1)(b4)(b7) Snapshots of number density distribution of C$^{6+}$ ions on the $y=0$ plane at times $t=93, 173$ and 320 fs, respectively. The magenta curves in (b1) plot the contour of laser intensity. (b2)(b5)(b8) are the same as (b1)(b4)(b7), but for longitudinal magnetic field $B_z$. (b3)(b6)(b9) are the same as (b1)(b4)(b7), but for azimuthal magnetic field $B_y$. (\textbf{c}) Evolution of on-axis characteristic parameters. The magnetic field strength $|\bf{B}|$ (red, solid) and number density of C$^{6+}$ ions $n_i$ (blue, dash) are averaged along the axis between 3 $\mu$m $< z <$ 17 $\mu$m. The normalized laser intensity $I/I_{\rm max}$ (green, dotted) is recorded at the entrance of the simulation box. The seed magnetic field generated by the Inverse Faraday Effect is effectively amplified by the converging plasma flows. \label{fig_1}}
\end{figure*}

Direct generation of a GG-scale magnetic field requires currents exceeding the Alfv${\rm {\acute{e}}}$n current limit \cite{Alfven} by several orders of magnitude. These, however, can hardly be achieved. Therefore, ultra-strong magnetic fields are generally obtained by amplification of an embedded (seed) magnetic field - a process called dynamo. The dynamo is active in many astrophysical objects including the Earth and the Sun. The ultra-strong magnetic fields occur spontanteously in the Universe at certain white dwarfs \cite{Garcia} and neutron stars \cite{Thompson}. In all these objects, a seed magnetic field is amplified to large scales and maintained by convection of electrically conducting fluids \cite{Moffatt}. However, amplification of magnetic field in laboratory is not straightforward. It is usually achieved by utilizing flux compression techniques. In this method, a field generated by traditional capacitors acts as a seed within a hollow conductor. The latter is squeezed inward by external drivers, such as mechanical explosives \cite{Fowler,Bykov}, electromagnetic force \cite{Cnare,Takeyama,Nakamura}, several nanosecond lasers \cite{Gotchev,Yoneda}, and $Z$-pinch \cite{Wessel,Felber,Appartaim,Ivanov,Gomez}. The magnetic force lines are compressed and the field is thus amplified. However, the amplified field is only about tens of megagauss (MG) even at a considerable compression ratio. The highest record of 90 MG was achieved in magnetized liner inertial fusion \cite{Gomez}, still falling behind the GG regime. Moreover, these methods strongly rely on mega science facilities, such as high energy nanosecond lasers and $Z$ machines, which are limited to few national laboratories. The ability to amplify magnetic field to the GG regime in smaller facilities is therefore of great interest for fundamental studies and applications.

In this Article, we propose an integrated scheme for the generation and amplification of magnetic field by irradiating a micro-channel target by a CP relativistic femtosecond laser. The Inverse Faraday Effect (IFE) caused by the circular laser polarization generates a seed longitudinal magnetic field in the target channel \cite{Najmudin,Liseykina}. After the laser-plasma interaction, the thermal expansion of the channel wall provides the electromotive force that drives the radial plasma flows. These converging inward flows switch on a plasma dynamo and amplify the seed magnetic field on the ion time scale within hundreds of femtosecond. Three-dimensional (3D) particle-in-cell (PIC) simulations show the peak strength of the amplified field approaches the GG level, which then persists on a picosecond time scale. Our results open the prospect of amplification of magnetic field to extreme strengths using readily available table-top Joule-class lasers.

\section*{Results}
\textbf{Converging plasma flows and amplification of seed magnetic field.} As illustrated in Fig. \ref{fig_1}(a), we use a micro-channel target as the plasma source. For definitiveness, in the simulation, the target length is $l=16$ $\mu$m and edge width is 8 $\mu$m. The channel is vacuum hollow inside with the internal radius of $R=2$ $\mu$m. The channel wall is fully ionized carbon plasma with the electron density $n_{e}=180n_{c}$ and C$^{6+}$ ion density $n_{i}=30n_{c}$. Here $n_{c}\sim1.1\times10^{21}{\rm {cm^{-3}}/\lambda_{L,\mu m}^{2}}$ is the critical density, and $\lambda_{L,\mu m}=0.8$ $\mu$m is the incident laser wavelength. For simplicity, the CP laser pulse is considered as a plane wave with the peak intensity $5\times10^{20}$ W/cm$^{2}$, the Gaussian temporal envelope of $1/e$ duration $\tau=30$ fs. The total laser energy in the simulation box corresponds to $W=12$ J.

\begin{figure}
	\centering
	\includegraphics[width=8.6cm]{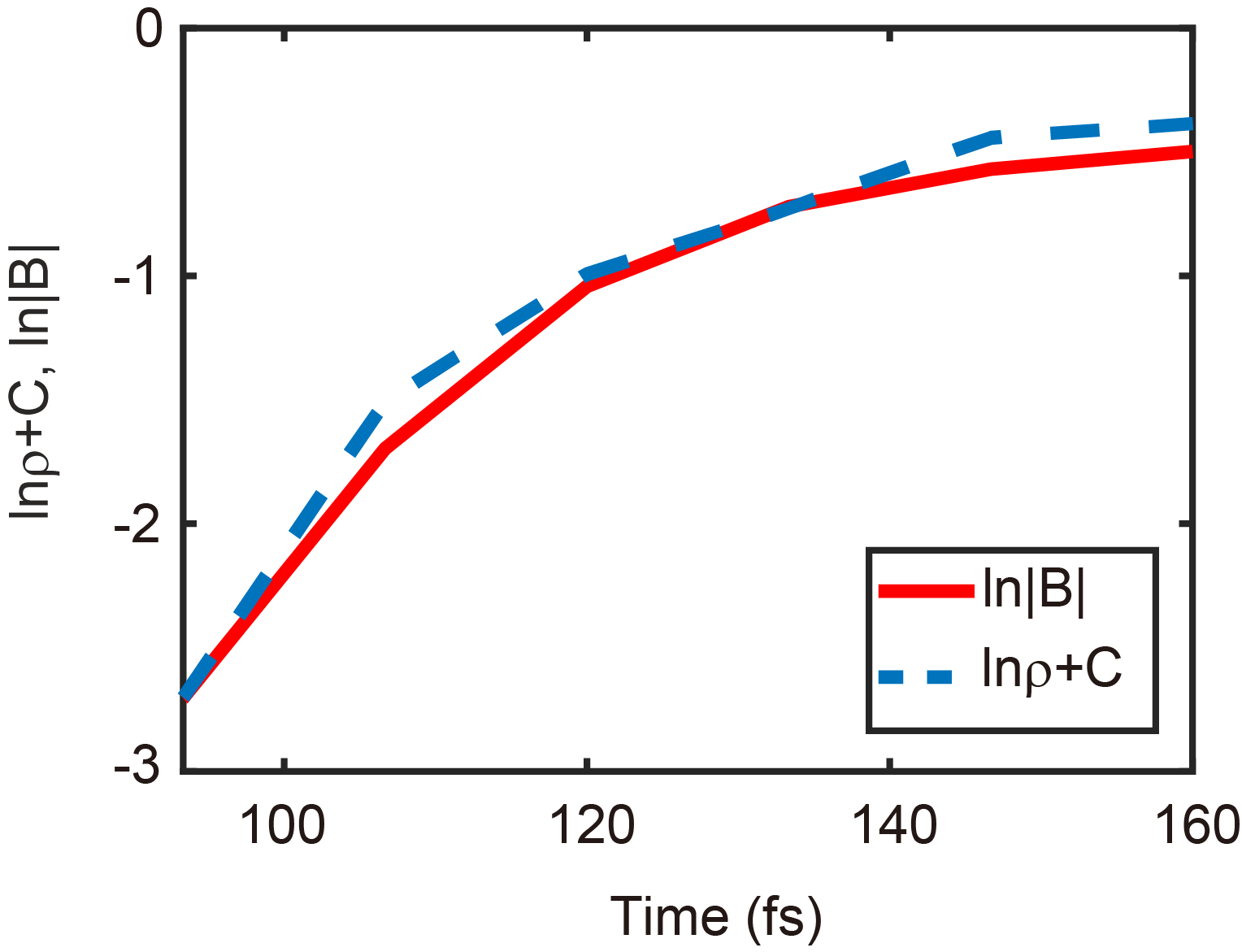}
	\caption{{\bf Evolution of on-axis $(\ln \rho + C)$ and $\ln |{\bf B}|$ during the amplification stage.} Here $C=1.4$ is a constant and $\rho$ is the plasma mass density. Both $\rho$ and $|{\bf B}|$ are averaged along the axis between 3 $\mu$m $< z <$ 17 $\mu$m. \label{fig_2}}
\end{figure}

The evolution of the plasma and magnetic field is summarized in Fig. \ref{fig_1}(b). The whole process consists of (I) seed magnetic field generation by IFE in the laser-plasma interaction period (Figs. \ref{fig_1} (b1)-(b3)), (II) field amplification (Figs. \ref{fig_1} (b4)-(b6)) and (III) maintenance (Figs. \ref{fig_1} (b7)-(b9)) by plasma dynamos from thermal expansion off the channel wall. Snapshots of distributions of C$^{6+}$ ion density $n_{i}$, longitudinal magnetic field $B_{z}$ and azimuthal magnetic field $B_{\phi}$ in each stage are presented to capture the characteristics of the process.

\begin{figure*}
	\centering
	\includegraphics[width=17.2cm]{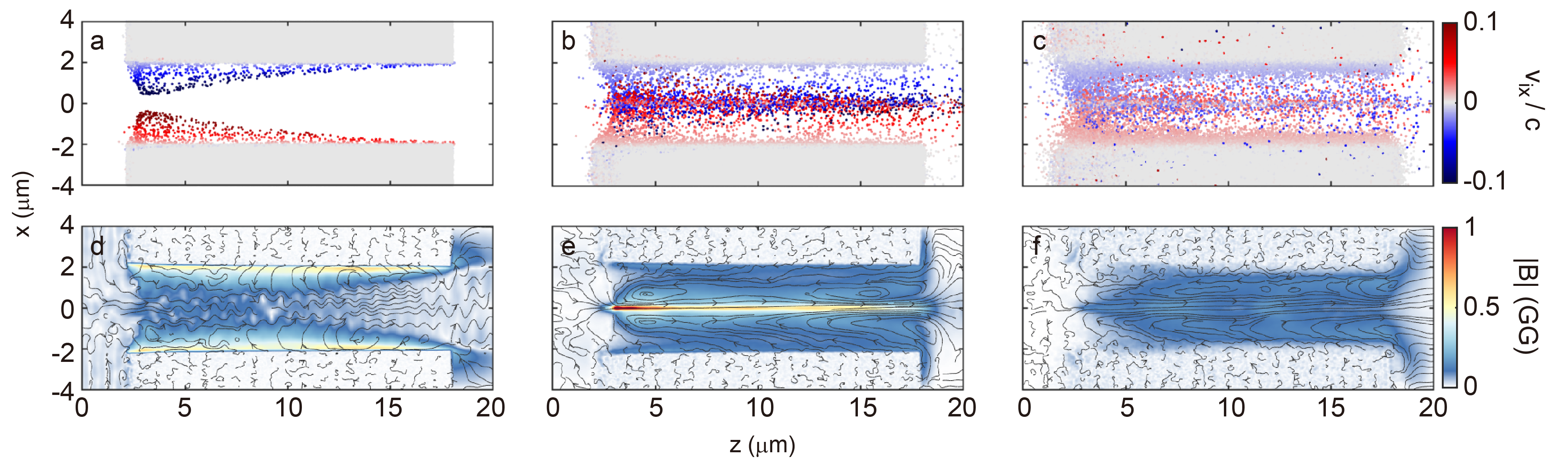}
	\caption{{\bf Details of the magnetic field amplification.} ({\bf a})-({\bf c}) Distributions of randomly selected 1\% of the total C$^{6+}$ ions on the $y=0$ plane at times $t=93, 173$ and 320 fs, respectively. The colors show normalized radial velocity $v_{ix}$ of C$^{6+}$ ions. ({\bf d})-({\bf f}) Distributions of magnetic field strength $|\bf{B}|$ and force lines on the $y=0$ plane at times $t=93, 173$ and 320 fs, respectively. \label{fig_3}}
\end{figure*}

In stage (I) of the seed field generation, the laser pulse (indicated by the magenta curves in Fig. \ref{fig_1}(b1)) interacts with the target. Electrons at the skin layers $l_{s}$ of the channel wall are pulled out by the laser electric field and accelerated preferentially towards the laser propagation direction by the ${\bf v}_{e}\times{\bf B}_{{\rm laser}}$ force \cite{Kaymak}. Here ${\bf v}_{e}$ is the electron velocity and ${\bf B}_{{\rm laser}}$ is the laser magnetic field. The forward-moving electrons induce a strong azimuthal magnetic field $B_{\phi}$ in the channel as shown in Fig. \ref{fig_1}(b3). Due to the circular laser polarization, each photon in such a laser carries the angular momentum $\hbar$. Accordingly, the total angular momentum of the laser pulse is $L_{{\rm laser}}=W/\omega$ \cite{Allan}. Here, $\omega$ is the laser frequency. During the interaction, the laser angular momentum $L_{{\rm laser}}$ is transferred to electrons \cite{Ju}. It results in a strong electron vortex and associated azimuthal current $J_{\phi}$. In our geometry, $l_{s}\ll R$ and $R\ll l$. The electrons that form the azimuthal current carry the angular momentum $L_{e}=m_{e}v_{e\phi}2\pi R^{2}l_{s}ln_{e}=2\pi m_{e}J_{\phi}R^{2}/e$. Here, $m_{e}$ is the electron mass and $e$ is the elementary charge. Assuming the conversion efficiency of angular momentum from laser to the azimuthal current is $\chi$, we obtain $J_{\phi}=\chi eW/2\pi m_{e}R^{2}\omega$. The seed magnetic field induced by IFE in the channel is then obtained from the Ampère's law $\nabla\times\mathbf{B}=4\pi\mathbf{j}/c$ \cite{Jackson}. Here, ${\bf j}$ is the current density and $c$ is the speed of light in vacuum. Integrating this equation, we get for the on-axis longitudinal magnetic field 
\begin{equation}
\frac{eB_{z}}{m_{e}c\omega}=2\chi\frac{r_{e}^{2}W}{k^{2}e^{2}R^{2}l},\label{eq1}
\end{equation}
where $r_{e}=e^{2}/m_{e}c^{2}$ is the classical electron radius and $k=\omega/c$ is the laser wave number. Given $\chi\sim1.3\times10^{-3}$ obtained from the simulation, the estimated $B_{z}\sim0.04$ GG, which agrees well with simulation results shown in Fig. \ref{fig_1}(b2). The seed magnetic field here is already at the same order of magnitude as the compressed field obtained in the OMEGA laser facility, where 40-beams of nanosecond lasers with on-target energy of 14 kJ were used \cite{Gotchev}.

Meanwhile, C$^{6+}$ ions at the skin layers of the channel wall move inward, forming converging plasma flows as a result of local charge imbalance, as shown in Fig. \ref{fig_1} (b1). We will  discuss later how such flows are vital for the stage (II) of magnetic field amplification. At $t=173$ fs, one sees in Fig. \ref{fig_1}(b4) that the flows meet on the channel axis and form a straight on-axis plasma density spike. The positive $B_{z}$, shown in Fig. \ref{fig_1}(b5), has formed a similar filament. Its radial profile is Gaussian-like, with FWHM of $R_{B_{z}}=0.3$ $\mu$m and a peak on-axis value larger than 0.6 GG. The product of $B_{z}R_{B_{z}}$ approaches the ignition threshold of $\simeq6\times10^{5}$ Gcm for the magneto-inertial fusion in cylindrical geometry \cite{Basko}. Thus our scheme - when scaled - becomes potentially useful for magnetized inertial confinement fusion. The distribution of $B_{\phi}$ is presented in Fig. \ref{fig_1}(b6). One sees that $B_{\phi}$ remains zero on axis due to the radial symmetry. This polarity of $B_{\phi}$ focuses relativistic electrons propagating in the positive $z$-direction. It suggests our scheme can be useful for charged beam collimation.

As the plasma flows cross the axis, the growth rate of the magnetic field seizes. Starting from the point when more plasma leaves the axis than moves towards it, the on-axis magnetic field and C$^{6+}$ ion density decrease, see Figs. \ref{fig_1} (b7)-(b9). Later at the stage (III), the on-axis C$^{6+}$ ion density is compensated by the thermal expansion of the bulk channel wall. The expansion provides electromotive force to maintain the magnetic field, as we can see the on-axis strength stays at about 100 MG on picosecond timescale.

\begin{figure}
	\centering
	\includegraphics[width=7.5cm]{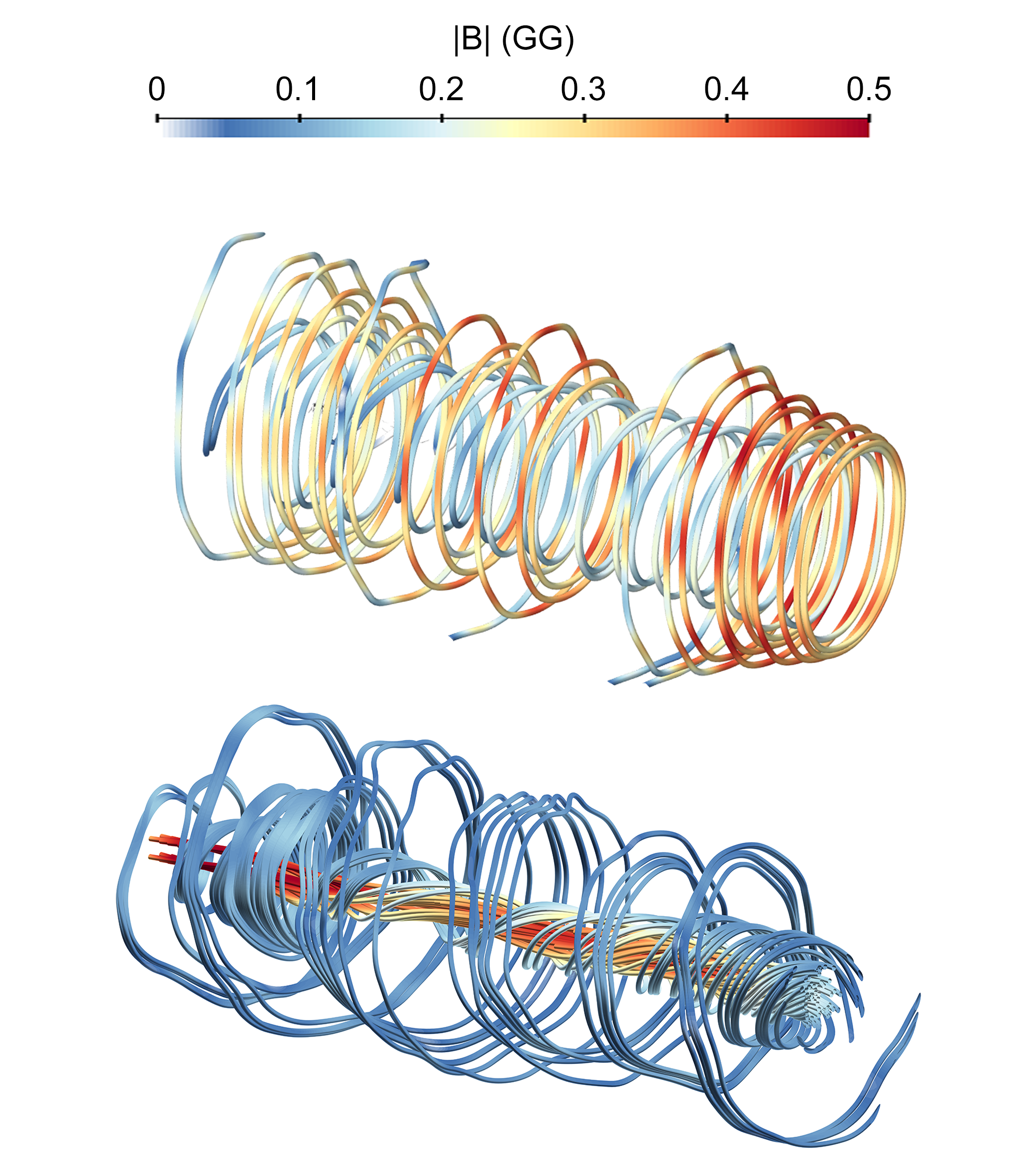}
	\caption{{\bf 3D perspective views of magnetic force lines at different times.} The upper and lower panel are at times $t=93$ and 173 fs, respectively. The colors show local strength of the field. \label{fig_4}}
\end{figure}

Figure \ref{fig_1}(c) presents the evolution of the on-axis magnetic field, C$^{6+}$ ion density as well as laser intensity. It further shows that the seed magnetic field grows synchronously with the ion density after the laser peak has passed through the target. The magnetic field peaks at 0.6 GG at 160 fs. It has been amplified up to 10 times within 70 fs. The peak value approaches that of dip poles of white dwarfs \cite{Wickramashinghe} and neutron stars \cite{Phinney}. The on-axis magnetic field stays at tens of MG after the fast drop between 173 fs to 300 fs, and then persists on sub-picosecond timescales. We see both the amplification and maintenance of the magnetic field occur after the laser-plasma interaction, and no additional drivers are required. The thermal expansion of the channel wall is on the ion time scale and the magnetic field can potentially persist for nanoseconds if we scale up the plasma channel size.

\begin{figure*}
	\centering
	\includegraphics[width=17.2cm]{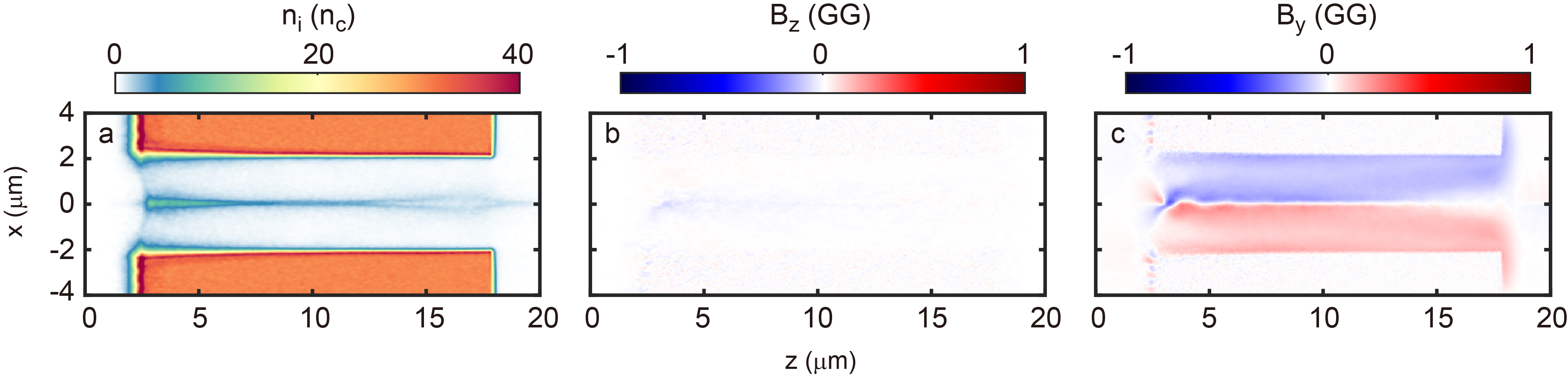}
	\caption{{\bf 3D-PIC simulation results for linearly polarized laser.} ({\bf a})-({\bf c}) Distributions of C$^{6+}$ ion number density, longitudinal magnetic field $B_z$, and azimuthal magnetic field $B_y$ on the $y=0$ plane at $t=173$ fs. \label{fig_5}}
\end{figure*}

\textbf{Mechanism for the magnetic field amplification.} To understand the mechanism for the amplification process, we consider the induction equation of magnetic field \cite{Chen}
\begin{equation}
\frac{\partial{\bf B}}{\partial t}=\nabla\times({\bf u}\times{\bf B})+\eta\nabla^{2}{\bf B},
\end{equation}
where \textbf{u} is fluid velocity and $\eta$ is an effective magnetic diffusivity defined in our simulation by small scale plasma turbulence. Due to axial symmetry, all fields depend on the radius $r$ only. The magnetic field has both the $\phi$- and $z$-component: ${\bf B}=(0,B_{\phi},B_{z})$ and the fluid velocity has radial component only: ${\bf u}=(u_{r},0,0)$. $\eta$ is relatively small and we neglect it for simplicity. One obtains in the cylinder coordinates

\begin{equation}
\frac{\partial B_{z}}{\partial t}=-\frac{1}{r}\frac{\partial}{\partial r}(ru_{r}B_{z}),\label{eq:Bz}
\end{equation}
and 
\begin{equation}
\frac{\partial B_{\phi}}{\partial t}=-\frac{\partial}{\partial r}(u_{r}B_{\phi}).\label{eq:Bphi}
\end{equation}

Eqs. \eqref{eq:Bz}-\eqref{eq:Bphi} are the dynamo equations that describe the magnetic field amplification. It is immediately clear that the seed magnetic field grows exponentially for converging plasma flows, i.e. $u_{r}<0$. Furthermore, by introducing the continuity equation for conservation of density, i.e. $\partial\rho/\partial t+\nabla\cdot(\rho{\bf u})=0$, one obtains the relation between on-axis $B_{z}$ and $\rho$

\begin{equation}
\ln B_{z}=\ln\rho+\kappa.
\end{equation}

Here $\rho$ is the mass density and $\kappa=1.4$ is a constant defined by the initial condition of the amplification. As shown in Fig. \ref{fig_2}, our theory agrees well with the simulation result during the amplification. In addition, as shown in Figs. \ref{fig_1}(b4)(b5), the amplified $B_{z}$ is indeed located at the density spike. Meanwhile $B_{\phi}$ must be zero on-axis as a result of axial symmetry.

Since the converging flow velocity $u_{r}$ is decisive for the magnetic field amplification, it is of interest to look into how $u_{r}$ evolves with time. Considering the C$^{6+}$ ions account more than 99\% of total fluid mass, we thus plot the C$^{6+}$ velocity distribution in Figs. \ref{fig_3} (a)-(c) to present the fluid velocity $u_{r}$. One sees the radial velocity of the ion flow front is at about 0.1$c$ during the laser-plasma interaction. The magnetic field grows as the ion flow converges. The on-axis field strength at the channel tip is found to be several times larger than that at the channel end, as a result of earlier formation of the converging flows. When the flows meet on the axis, $u_{r}$ must be reconsidered, as shown in Fig. \ref{fig_3} (b). $|u_{r}|$ decreases as well as the growth rate of the magnetic field until the field reaches its peak strength when $u_{r}=0$. Comparing Fig. \ref{fig_3}(d) and (e), one sees how the magnetic force lines are organized from the initial wavy curves into straight lines on the axis during the amplification stage.

\begin{table*}
	\caption{\label{tb1} Summary of the target material influence on the magnetic field amplification.}
	\setlength{\tabcolsep}{3mm}{
		\begin{tabular}{cccc} \toprule
			Material  & Ion charge-to-mass ratio ($e/m_H$) &  Peak value (GG)  &  Peak time (ps)   \\ 
			\hline
			Hydrogen & 1 & 0.86 & 0.14    \\
			CH & 0.5 \& 1 & 0.65 & 0.17  \\
			Carbon & 0.5 & 0.61 & 0.16  \\
			DT & 0.5 \& 0.33 & 0.48 & 0.16  \\
			\hline
	\end{tabular}}
\end{table*}

The detailed 3D perspective views of the magnetic force lines before and after the amplification are illustrated in Fig. \ref{fig_4}. The initial longitudinal seed field is much weaker than the azimuthal one. The force lines appear to be spirals with the peak strength close to the channel wall. The longitudinal magnetic field accounts only for 1.5\% of the total magnetic field energy. After the amplification, one clearly sees the change of the topology of magnetic force lines. The initial loose azimuthal magnetic force lines wind into tight and twisted bunches on the axis. The on-axis magnetic field is 5 times larger than that at the channel periphery. The energy deposited in the longitudinal magnetic field grows up to 15.5\% of the total magnetic field energy.

The fluid velocity $u_{r}$ turns into positive from $t=173$ to 320 fs, due to the fact that more plasma leaves the axis than moves towards it. Therefore the magnetic field decreases. Later, $u_{r}$ is again compensated by the thermal expansion of the target wall, as indicated in Fig. \ref{fig_3}(c) by the large red and blue areas adjacent to the target inner surface. The magnetic field is stabilized and stays at the 100 MG level for a very long time. The magnetic force lines still retain the well-defined straight shape on the axis, as shown in Fig. \ref{fig_3}(f).

\textbf{Role of laser polarization and target materials.} As indicated by Eq. \ref{eq:Bz}, the amplification takes place only when a seed magnetic field is embedded in the channel. In our previous case, the seed magnetic field is generated by IFE from laser polarization. Here we discuss briefly on the influence of the latter. Fig. \ref{fig_5} shows the results from a linearly polarized (LP) laser of the same input energy of 12 J, while the other parameters are identical with the previous CP case. We indeed observe the formation of an on-axis density spike (Fig. \ref{fig_5}(a)) and a similar distribution of the azimuthal magnetic field (Fig. \ref{fig_5}(c)). The formation of the on-axis density spike indicates that in both LP and CP case, the walls undergo similar expansion, which provides the electromotive force that drives radial flows. However, since LP laser does not carry any angular momentum, the electron vortex for the seed magnetic field generation is absent. As a result, neither the seed longitudinal magnetic field, nor its amplification is observed, as shown in Fig. \ref{fig_5}(b). 

The robustness of our scheme is demonstrated by several simulations with different target materials. The main results are summarized in Table. \ref{tb1}. In all the cases, ions are considered to be fully ionized. We observe the three stages (i.e. generation, amplification, and maintenance) of magnetic field evolution in all the simulations. The peak strength of amplified magnetic field is found to be positively correlated with $\sqrt{q/m_{i}}$, where $q$ is the ion charge. It can be interpreted that larger $\sqrt{q/m_{i}}$ leads to larger $u_{r}$, and accordingly stronger peak magnetic field. By comparing the hydrogen and carbon case, one can conclude that larger $\sqrt{q/m_{i}}$ also results in shorter amplification time. However, it becomes complicated with multi-species targets, since the light ions also contribute to the field amplification. The detailed influence of target materials is beyond the scope of this Article, and will be presented elsewhere.

\section*{Discussion}

\begin{figure}
	\centering
	\includegraphics[width=8.6cm]{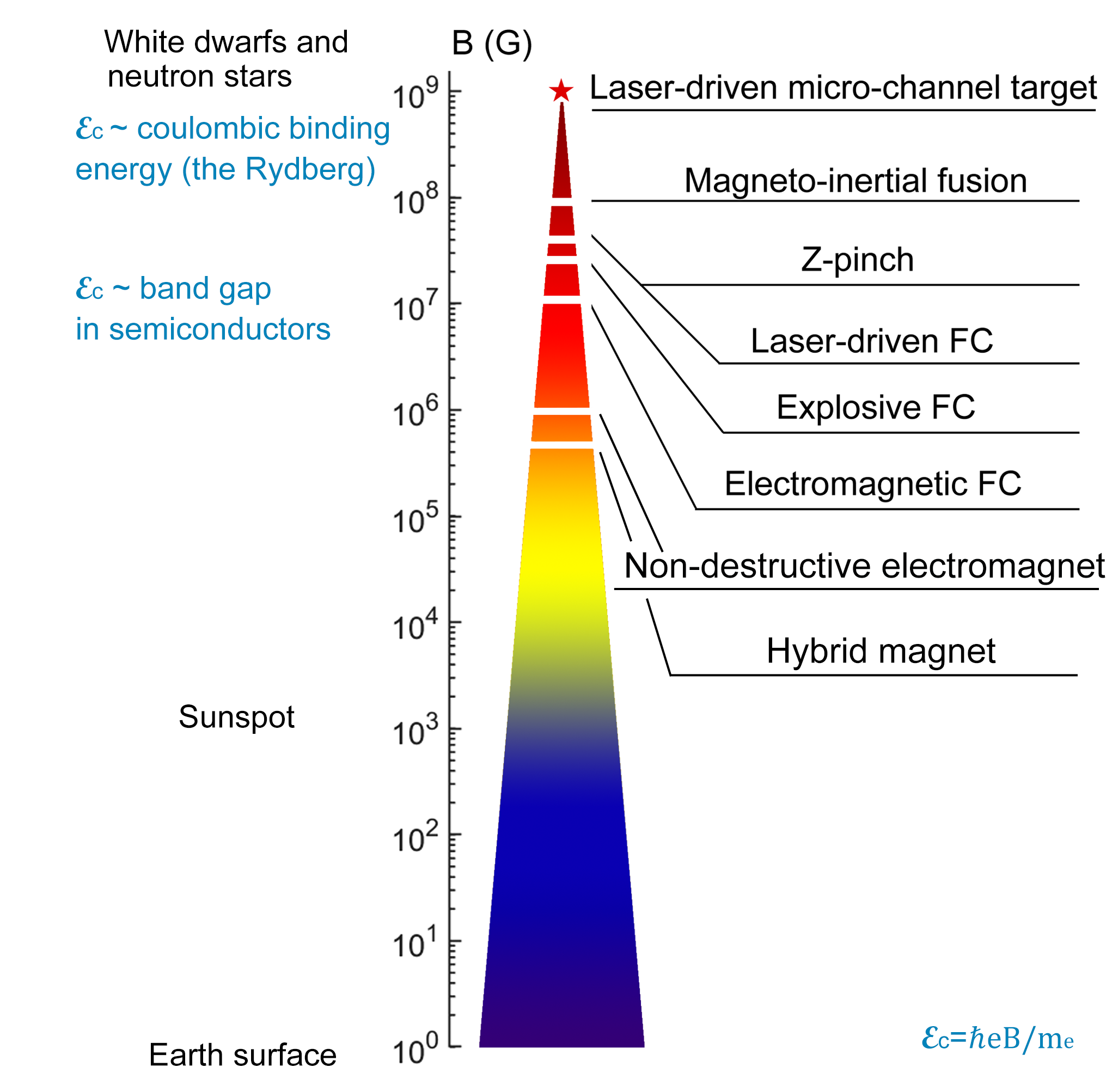}
	\caption{{\bf Parameter space of magnetic field strength showing the regime accessible by irradiation of micro-channel targets relative to other physical environments.} Typical field strengths achieved by other methods are shown on the right. \label{fig_6}}
\end{figure}

It is worth mentioning that the GG-scale magnetic field can be potentially produced by intense laser-plasma interaction. It has been reported that fields at hundreds of MG or higher can be found in laser-driven plasma channels \cite{Jiang}, at the rare side of laser-heated plain targets \cite{Nakatsutsumi}, or adjacent to filamented currents \cite{Mondal}. In those cases, the produced fields are usually azimuthal, due to electrons preferentially moving along the laser propagation direction. These are poorly suitable for applications such as magneto-inertial fusion. Nevertheless, there are two special geometries for strong longitudinal magnetic field generation, i.e. laser ablation on capacitor-coil targets \cite{Fujioka} and laser irradiation at helical targets \cite{Jiang1,Korneev}. In both cases, the field is generated by currents originated from local charge imbalance, which is completely different from the dynamo amplification in our scheme.

Fig. \ref{fig_6} illustrates the parameter space of magnetic field strength. At the magnetic field of 10 MG, the electron cyclotron energy $\mathcal{E}_{c}$ is of the same order of magnitude as a typical band gap of semiconductors. The Bloch electron model, based on the assumption of atomic periodic potential, no longer holds. It leads to abundant novel effects in electronic properties of materials. Such a strong magnetic field has been achieved by several different approaches. The field produced in our scheme reaches a completely new level of GG scale. This is nearly one order of magnitude larger than the one demonstrated in magneto-inertial fusion experiments. Our scheme sets a possible platform for reproducing extreme magnetic environment in laboratories using readily available table-top Joule-class femtosecond lasers.

\section*{Methods}
The 3D PIC simulations presented in this work are conducted with EPOCH code \cite{Arber}. A CP plane wave with wavelength $\lambda=800$ nm, peak intensity $5\times 10^{20}$ W/cm$^2$, and Gaussian temporal envelope of $1/e$ duration 30 fs is used to simulate the laser pulse. Periodic lateral boundaries are used for both particles and fields. Thus, our simulations correspond to an array of plasma channels. The single channel is of longitudinal length 16 $\mu$m and edge length 8 $\mu$m, with vacuum hollow inside of an internal radius 2 $\mu$m. The channel wall is fully ionized carbon plasma with electron density of $180n_c$  and C$^{6+}$ ion density of $30n_c$ from $r=2$ to 4 $\mu$m. Here $n_c \sim 1.1 \times 10^{21} \rm{cm^{-3}}/\lambda^2$ is the critical density. The dimensions of the simulation box are $x \times y \times z = 8$ $\mu$m $\times 8$ $\mu$m $\times 20$ $\mu$m, which is sampled by $600 \times 600 \times 1000$ cells with 4 macro particles for both electrons and carbons. 

The results shown in Table. \ref*{tb1} are obtained using identical simulation setups, except for different ion species. In all the cases, ions are considered to be fully ionized.

Our main results presented in this Article are also checked with virtual laser-plasma laboratory (VLPL) code \cite{Pukhov,Pukhov1}. The simulation box is sampled by $160\times 160 \times 1300$ cells with 8 macro particles for both electrons are carbons.

\section*{Acknowledgment}
This work is supported by the DFG (project PU 213/9), GCS Juelich (project QED20), the National Key R\&D Program (Grant No. 2016YFA0401100), the National Natural Science Foundation of China (Grant Nos. 11575031, 11705120, U1630246 and 11875092), the Natural Science Foundation of Top Talent of SZTU (grant no. 2019010801001). K.J. acknowledges the support from China Scholarship Council. K.J. would like to thank X. F. Shen, C. Baumann, V. Kaymak, G. Lehmann, L. Reichwein and T. Y. Long for useful discussions and help.

\section*{Author contributions}
K.J. and A.P. developed the theoretical work. A.P. developed the VLPL code. K.J. and A.P. conducted the simulations. K.J. analyzed the data and produced the figures. K.J. and A.P. wrote the Article. All of the authors discussed the results and commented on the Article.

\section*{Data availability}
The data used in this work are available from the corresponding authors upon request.

\section*{Competing financial interests}
The authors declare no competing financial interests.
\end{document}